\documentclass[twocolumn]{aastex62}

\graphicspath{{./}{figures/}}

\submitjournal{ApJ}

\shorttitle{Pulsars in GC M13}
\shortauthors{L Wang et al.}

\usepackage{afterpage}
\usepackage{lipsum}
\usepackage{rotating}
\usepackage{amsmath}
\usepackage{graphics}

\begin{document}

\title{Discovery and timing of pulsars in the globular cluster M13 with FAST}

\correspondingauthor{Lin Wang; B.\,W.\,Stappers }
\email{wanglin@bao.ac.cn; Ben.Stappers@manchester.ac.uk}

\author{Lin Wang}
\affil{CAS Key Laboratory of FAST, National Astronomical Observatories, Chinese Academy of Sciences \\
Beijing 100101, P.\,R.\,China}
\affiliation{Jodrell Bank Centre for Astrophysics, School of Physics and Astronomy, The University of Manchester\\
Manchester, M13 9PL, UK}
\affiliation{School of Astronomy and Space Science, University of Chinese Academy of Sciences\\
Beijing, P.\,R.\,China}

\author{Bo Peng}
\affiliation{CAS Key Laboratory of FAST, National Astronomical Observatories, Chinese Academy of Sciences \\ Beijing 100101, P.\,R.\,China}

\author{B.\,W.\,Stappers}
\affiliation{Jodrell Bank Centre for Astrophysics, School of Physics and Astronomy, The University of Manchester\\
Manchester, M13 9PL, UK}

\author{Kuo Liu}
\affiliation{Max-Plank-Institut f{\"u}r Radioastronomie, Auf dem H{\"u}gel 69, \\
Bonn, D-53121 , Germany}
\affiliation{CAS Key Laboratory of FAST, National Astronomical Observatories, Chinese Academy of Sciences \\
Beijing 100101, P.\,R.\,China}

\author{M.\,J.\,Keith}
\affiliation{Jodrell Bank Centre for Astrophysics, School of Physics and Astronomy, The University of Manchester\\
Manchester, M13 9PL, UK}

\author{A.\,G.\,Lyne}
\affiliation{Jodrell Bank Centre for Astrophysics, School of Physics and Astronomy, The University of Manchester\\
Manchester, M13 9PL, UK}

\author{Jiguang Lu}
\affiliation{CAS Key Laboratory of FAST, National Astronomical Observatories, Chinese Academy of Sciences \\ Beijing 100101, P.\,R.\,China}

\author{Ye-Zhao Yu}
\affiliation{Qiannan Normal University for Nationalities \\ Duyun 558000, P.\,R.\,China}

\author{Feifei Kou}
\affiliation{CAS Key Laboratory of FAST, National Astronomical Observatories, Chinese Academy of Sciences \\ Beijing 100101, P.\,R.\,China}
\affiliation{Xinjiang Astronomical Observatories, Chinese Academy of Sciences\\ Urumqi 830011,P.\,R.\,China}

\author{Jun Yan}
\affiliation{CAS Key Laboratory of FAST, National Astronomical Observatories, Chinese Academy of Sciences \\ Beijing 100101, P.\,R.\,China}

\author{Peng Jiang}
\affiliation{CAS Key Laboratory of FAST, National Astronomical Observatories, Chinese Academy of Sciences \\ Beijing 100101, P.\,R.\,China}

\author{Chengjin Jin}
\affiliation{CAS Key Laboratory of FAST, National Astronomical Observatories, Chinese Academy of Sciences \\ Beijing 100101, P.\,R.\,China}

\author{Di Li}
\affiliation{CAS Key Laboratory of FAST, National Astronomical Observatories, Chinese Academy of Sciences \\ Beijing 100101, P.\,R.\,China}

\author{Qi Li}
\affiliation{CAS Key Laboratory of FAST, National Astronomical Observatories, Chinese Academy of Sciences \\ Beijing 100101, P.\,R.\,China}

\author{Lei Qian}
\affiliation{CAS Key Laboratory of FAST, National Astronomical Observatories, Chinese Academy of Sciences \\ Beijing 100101, P.\,R.\,China}

\author{Qiming Wang}
\affiliation{CAS Key Laboratory of FAST, National Astronomical Observatories, Chinese Academy of Sciences \\ Beijing 100101, P.\,R.\,China}

\author{Youling Yue}
\affiliation{CAS Key Laboratory of FAST, National Astronomical Observatories, Chinese Academy of Sciences \\ Beijing 100101, P.\,R.\,China}

\author{Haiyan Zhang}
\affiliation{CAS Key Laboratory of FAST, National Astronomical Observatories, Chinese Academy of Sciences \\ Beijing 100101, P.\,R.\,China}

\author{Shuxin Zhang}
\affiliation{CAS Key Laboratory of FAST, National Astronomical Observatories, Chinese Academy of Sciences \\ Beijing 100101, P.\,R.\,China}

\author{Yan Zhu}
\affiliation{CAS Key Laboratory of FAST, National Astronomical Observatories, Chinese Academy of Sciences \\ Beijing 100101, P.\,R.\,China}


\collaboration{(The FAST collaboration)}



\begin{abstract}

We report the discovery of a binary millisecond pulsar (namely PSR J1641+3627F or M13F) in the globular cluster M13 (NGC 6205) and timing solutions of M13A to F using observations made with the Five-hundred-metre Aperture Spherical radio Telescope (FAST). PSR J1641+3627F has a spin period of 3.00 ms and an orbital period of 1.4 \,days. The most likely companion mass is 0.16 M$_{\odot}$.  M13A to E all have short spin periods and small period derivatives. We also confirm that the binary millisecond pulsar PSR J1641$+$3627E (also M13E) is a black widow with a companion mass around 0.02 M$_{\odot}$. We find that all the binary systems have low eccentricities compared to those typical for globular cluster pulsars and that they decrease with distance from the cluster core.  This is consistent with what is expected as this cluster has a very low encounter rate per binary. 


\end{abstract}

\keywords{globular cluster: individual (M13) --- pulsars: individual (J1641+3627F) }


\section{Introduction} \label{sec:intro}
Since the first globular cluster (GC) pulsar PSR B1821$-$24A was discovered in M28 by \citet{Lyne1987}, the total number of pulsars currently known in GCs is at least 155\footnote{A catalog of GC pulsars is maintained by P. C. Freire at http://www.naic.edu/~pfreire/GCpsr.html}. GCs are known as efficient factories for the formation of millisecond pulsars (MSPs). GCs contain a large number of low-mass X-ray binaries (LMXBs) \citep{Clark1975}, which produce MSPs through accretion \citep{Camilo2005}. The large stellar densities in GCs provide ideal environments for stellar interactions \citep{Camilo2005}， which increase the possibility of forming binaries and MSPs \citep[e.g.][]{Cadelano2018,Verbunt2014}.


M13 (NGC 6205) is a bright GC (V $\approx$ 5.8) located in the constellation of Hercules ($\alpha$ $=$ 16$^{h}$41$^{m}$41$\fs$24, $\delta$ $=$ +36$^\circ$27$^{'}$35$\farcs$5), with a distance of 7.1\,kpc \citep{Harris1996}. The GC core and half$-$light radius are 0$\farcm$62 and 1$\farcm$49, respectively. There are five known pulsars in M13 so far and they all have a dispersion measure (DM) around 30\,pc cm$^{-3}$.  M13A and M13B were discovered in an Arecibo GC survey at a centre frequency of 430\,MHz in 1991\citep{Kulkarni1991}. M13A is an isolated pulsar with a spin period of 10.37\,ms, the longest among the five known pulsars; M13B has a spin period of 3.53\,ms and is in a binary system with an orbital period of 1.3 days \citep{Kulkarni1991}. M13C, M13D and M13E were discovered in a 1.4-GHz Arecibo survey in 2007 \citep{Hessels2007}. M13C is an isolated pulsar with a spin period of 3.72\,ms, while M13D and M13E have spin periods of 3.12\,ms and 2.49\,ms, respectively. The latter two are both in binary systems, with orbital periods of 0.6 and 0.1\,days, respectively. M13E was also reported to possibly be a black widow system. \\

In this paper we present the discovery of the binary pulsar PSR J1641$+$3627F (M13F) and timing solutions of all the known pulsars in the GC M13. In the next section, we introduce FAST and our observing modes. Data analysis is presented in Section \ref{sec:analysis}. The discovery and timing solutions of the pulsars in the GC M13 are presented in Section \ref{sec:results}. Discussion and conclusion are made in Sections \ref{sec:discuss} and \ref{sec:conclusion }.\\

\section{Observations} \label{sec:obs}

The Five-hundred-metre Aperture Spherical radio Telescope (FAST) is located in the Karst depression region in Guizhou, China. It has a diameter of 500 meters and an illuminated aperture of 300 meters (\citet{Peng2001};\citet{Li2013}). 
We have used FAST to observe M13 since 2017, with both the Ultra-Wide-Band receiver (henceforth UWB) and the 19-beam receiver (henceforth 19-beam). M13 was chosen as the first target for our GC observation campaign since it can be tracked within a zenith angle of 26.4 degrees by FAST for a few hours, which leveraged the efficiency of the telescope and reduced the tension in the cable network holding the main reflector at the beginning of its commissioning. 
\\
The first observation of M13 was made in 2017 with the UWB, while the remaining twenty-four were conducted with the 19-beam. The UWB covers the frequency range of 270$-$1620\,MHz. For our observations, the wide-band signal was first filtered into two sub-bands with frequency range of 270$-$800\,MHz and 1200$-$1620\,MHz, respectively, to avoid frequencies with strong interference. Then the lower sub-band signal was sent to a Reconfigurable Open Architecture Computing Hardware (ROACH)\footnote{developed by the Collaboration for Astronomy Signal Processing and Electronics Research (CASPER) group; http://casper.berkeley.edu/} unit, where the signal were sampled at Nyquist frequency, digitised into 8-bit real samples with dual polarisation, packetized and stored on a hard-drive device. 
 The follow-up observations were undertaken using the 19-beam receiver which replaced the UWB in July 2018.
 It covers a frequency range of 1050$-$1450\,MHz with a system temperature below 20\,K on cold sky \citep{Jiang2019}. During the observations, the signal from the central beam was Nyquist sampled at 8-bit resolution and then a ROACH was used to divide the overall band into 4096 channels each 12\,kHz wide. Power in each of the two polarizations was averaged in time, which resulted in a 50\,$\mu$s time resolution. The data were then recorded in the PSRFITS format \citep{Hotan2004}. A summary of all observations used in this paper is given in Table \ref{tab:observations_table}, which includes the central frequencies, bandwidths, observation lengths, number of pulsars detected and the flux density upper limits.
\\

\startlongtable
\begin{deluxetable*}{c|cccccccr}
\tablecaption{Summary of observations made with FAST for M13 \label{tab:observations_table}}
\tablehead{
\colhead{Date} & \colhead{MJD} & \colhead{Receiver} & \colhead{$f_{\text{c}}$} & \colhead{BW} &  \colhead{$T_{\text{obs}}$} & \colhead{known PSR} & \colhead{F detected} & \colhead{flux limit} \\
\colhead{YYYY-mm-dd} & \colhead{} & \colhead{} & \colhead{(MHz)} & \colhead{(MHz)} & \colhead{(min)} & \colhead{} & \colhead{Y/N} & \colhead{($\mu$Jy)}
}
\startdata
        2017-12-20 & 58107 & UWB & 535 & 530 & 30 & AB & Y & \nodata \tablenotemark{b} \\
        2018-10-06 & 58397 & 19-beam & 1250 & 400 & 228 & ABCDE & Y & 0.47  \\
        2018-10-30 & 58421 & 19-beam & 1250 & 400 & 180 & ABCDE & Y & 0.53  \\
        2018-11-28 & 58450 & 19-beam & 1250 & 400 & 60 & ABCDE & N & 0.92  \\
        2018-12-02 & 58454 & 19-beam & 1250 & 400 & 60 & ABCDE & N & 0.92  \\
        2018-12-03 & 58455 & 19-beam & 1250 & 400 & 60 & ABCDE & N & 0.92  \\
        2018-12-05 & 58457 & 19-beam & 1250 & 400 & 60 & ABCDE & N & 0.92  \\
        2018-12-07 & 58459 & 19-beam & 1250 & 400 & 60 & ABCDE & Y & 0.92  \\
        2018-12-08 & 58460 & 19-beam & 1250 & 400 & 60 & ABCDE & Y & 0.92  \\
        2018-12-17 & 58469 & 19-beam & 1250 & 400 & 60 & ABCDE & Y & 0.92  \\
        2019-03-19 & 58545 & 19-beam & 1250 & 400 & 180 & ABCD & Y & 0.53  \\
        2019-03-20(1)\tablenotemark{a} & 58546 & 19-beam & 1250 & 400 & 60 & ABCD & Y & 0.92  \\
        2019-03-20(2)\tablenotemark{a} & 58546 & 19-beam & 1250 & 400 & 60 & ABCD & Y & 0.92  \\
        2019-03-21 & 58547 & 19-beam & 1250 & 400 & 60 & ABCDE & Y & 0.92  \\
        2019-03-23 & 58549 & 19-beam & 1250 & 400 & 60 & ABCE & Y & 0.92  \\
        2019-03-24 & 58550 & 19-beam & 1250 & 400 & 60 & ABCD & Y & 0.92  \\
        2019-03-29 & 58555 & 19-beam & 1250 & 400 & 60 & ABCE & Y & 0.92  \\
        2019-07-20 & 59684 & 19-beam & 1250 & 400 & 60 & ABCDE & Y & 0.92  \\
        2019-07-21 & 59685 & 19-beam & 1250 & 400 & 60 & ABCDE & Y & 0.92  \\
        2019-07-23 & 59687 & 19-beam & 1250 & 400 & 60 & ABCDE & Y & 0.92  \\
        2019-08-17 & 58712 & 19-beam & 1250 & 400 & 60 & ABCDE & N & 0.92 \\
        2019-08-28 & 58723 & 19-beam & 1250 & 400 & 60 & ABCDE & Y & 0.92 \\
        2019-08-31 & 58726 & 19-beam & 1250 & 400 & 60 & ABCDE & Y & 0.92 \\
        2019-09-06 & 58732 & 19-beam & 1250 & 400 & 60 & ABCDE & Y & 0.92 \\
        2019-09-14 & 58740 & 19-beam & 1250 & 400 & 90 & ABCDE & Y & 0.75 \\
\enddata
\tablenotetext{a}{There are two sessions on the same day.}
\tablenotetext{b}{The UWB data were acquired to test the overall observing system of the telescope at the beginning of the commissioning phase. Key parameters such as the reflector efficiency and the system temperature were not well understood, which makes it difficult to estimate the flux density limit.}

\end{deluxetable*}

\section{Data Analysis} \label{sec:analysis}
\subsection{Data Format}
\begin{figure}[ht!]
\epsscale{1.3}
\plotone{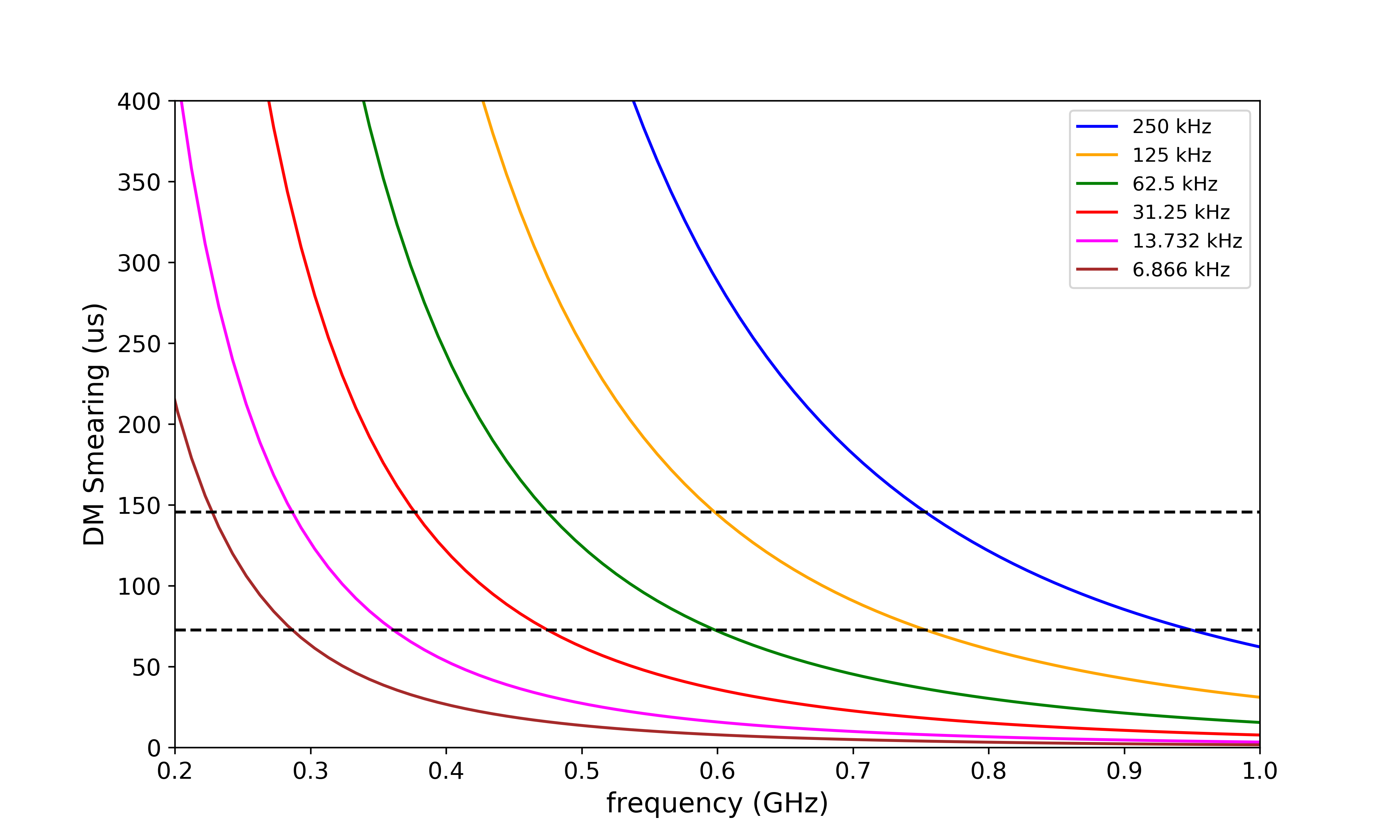}
\caption{DM Smearing as a function of frequency. The colored lines are different channel bandwidth corresponding different number of channels. The brown and purple lines correspond to the numbers of channels we used in the two sub-bands.   \label{tab:DM_smearing}}
\end{figure}

The voltage data from the UWB were first channelised, down-sampled, summed in polarization and written out in PSRFITS search-mode format\footnote{The conversion used the \textsc{UDP2psrfits} (developed by K.~Liu) routine which is available via: \url{https://github.com/xuanyuanstar/psrcov}.} \citep{Hotan2004}. During the conversion, the data were split into two sub-bands in order to decrease computing power and DM smearing within a channel. One frequency band spanned 250$-$400\,MHz \footnote{Although the low-frequency band of UWB covers a range of 270$-$800\,MHz, we saw signal that starts at 250\,MHz. Thus we kept the data from 250\,MHz} and the other spanned 400$-$800\,MHz. 
The numbers of channels are 21847 (with channel bandwidth of 6.9\,kHz) and 29128 (with channel bandwidth of 13.7\,kHz) for the two sub-bands respectively which were chosen by both minimising the dispersion smearing within a channel for the cluster DM of 30\,pc cm$^{-3}$ and equaling the DM smearing and sampling time.
These number of channels resulted in sampling times in the two sub-bands of 145.64\,$\mu$s and 72.82\,$\mu$s, respectively. Figure \ref{tab:DM_smearing} shows DM smearing within a channel as a function of frequency, the upper and lower dash line represent the sample times in the two sub-bands. We also estimated the effects of scattering using the empirical formula (t$_{\rm scatt}$ is in\,ms) \citep{Bhat2004}
\begin{equation}
\small
{\rm lg}(t_{\rm scatt}) = -6.46 + 0.154~{\rm lg}({\rm DM})  +
1.07~({\rm lg}({\rm DM}))^{2}
 -3.86~{\rm lg}\left(\frac{\nu}{{\rm GHz}}\right) 
\label{scatt.eqn}
\end{equation}

Even at 250\,MHz, the scattering effect is estimated to be 27\,$\mu$s for M13 which is much smaller than the sampling time. The data from the 19-beam receiver were recorded in the search mode PSRFITS format. As described in Section \ref{sec:obs}, it has 4096 channels and a sampling time of 50\,$\mu$s. The data covers a frequency range of 1000-1500\,MHz.  \\

\subsection{Search Techniques}

All the data were processed using the PRESTO software package \citep{Ransom2002}. We used the {\tt rfifind} package to generate a RFI (Radio Frequency Interference) mask in a similar way to that described in \citet{Hessels2007}. The mask was used with {\tt prepsubband} to generate 100 de-dispersed time series at a central DM $=$ 30\,pc cm$^{-3}$ and steps of 0.1\,pc cm$^{-3}$. The DM step size was chosen to have  small DM smearing caused by an incorrect DM and sensible requirements on processing. DM smearing caused by an incorrect DM in channels is given by 

\begin{equation}
\small
t_{\Delta{\rm DM}} = 4.1\left[\left(\frac{\nu_{\rm low}}{{\rm GHz}}\right)^{-2} -
\left(\frac{\nu_{\rm high}}{{\rm GHz}}\right)^{-2}\right] \left(\frac{\Delta{\rm
DM}}{{\rm pc~cm^{-3}}}\right){\rm ms},
\end{equation} 
where $\nu_{\rm low}$ and $\nu_{\rm high}$ are the lowest and highest edges of the frequency bandwidth. $\Delta{\rm DM}$ is the DM step size that we have chosen.
 In our cases, it introduces a maximum DM smearing within a channel of 0.4 $\mu$s which is much smaller than the sampling time. The DM range was determined by considering the DM difference between the pulsar with the largest and smallest DM in GCs with more than one pulsar. The largest difference is 9.93\,pc cm$^{-3}$, thus we use a DM range of 10\,pc cm$^{-3}$ centred around the average DM of the known pulsars in M13. \\
A fast Fourier transform was applied to each of these de-dispersed time series to obtain the power spectra and red noise was removed from each power spectra using the default settings for {\tt rednoise}. After de-dispersion, the effects from the rotation of the Earth and its motion around the sun were removed.

Finally, an acceleration search was performed using {\tt accelsearch} on the spectra with an acceleration range specified with the parameter zmax = 300. This parameter is the largest drift of the pulse frequency across frequency bins caused by acceleration and an explanation can be found in \citet{Ransom2002}. We also searched each data set after dividing them into 20 minute sub-integrations in order to be sensitive to systems with minimum orbital periods of 3 hours. The initial candidates were sifted using the code \textsc{sift}, which is part of the pulsarTool\footnote{\url{https://github.com/mitchmickaliger/pulsarTools}} software package. Each candidate was searched for periods that have harmonic ratios from 1 to 16. The initial parameters resulting from the search  for the candidates were used to re-fold the raw search mode data and the results were visual inspected.  \\

\section{Results} \label{sec:results}


\begin{figure*}[ht!]
\plotone{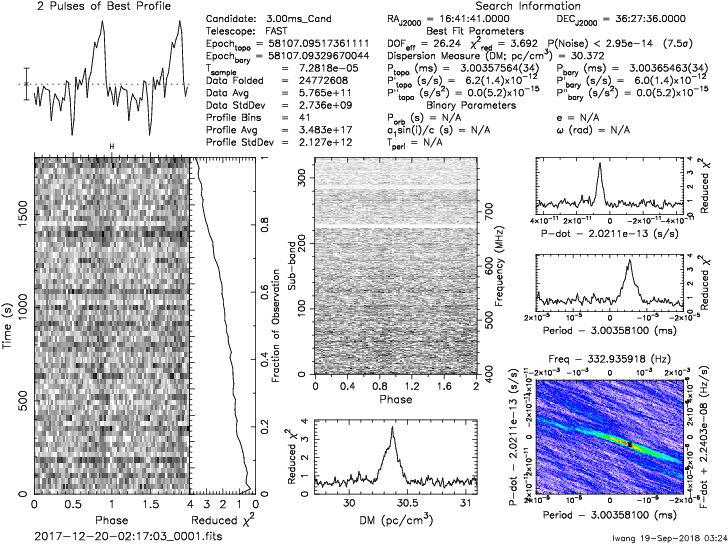}
\caption{PRESTO discovery plot of M13F in the 400$-$800 MHz sub-band of the UWB. The observation was made on December 2017 with an integration time of 30 minutes. No signal was detected in the simultaneously observed 250$-$400 MHz sub-band.  \label{tab:searching_result}}
\end{figure*}

\begin{figure*}[ht!]
\centering
\epsscale{1.3}
\plotone{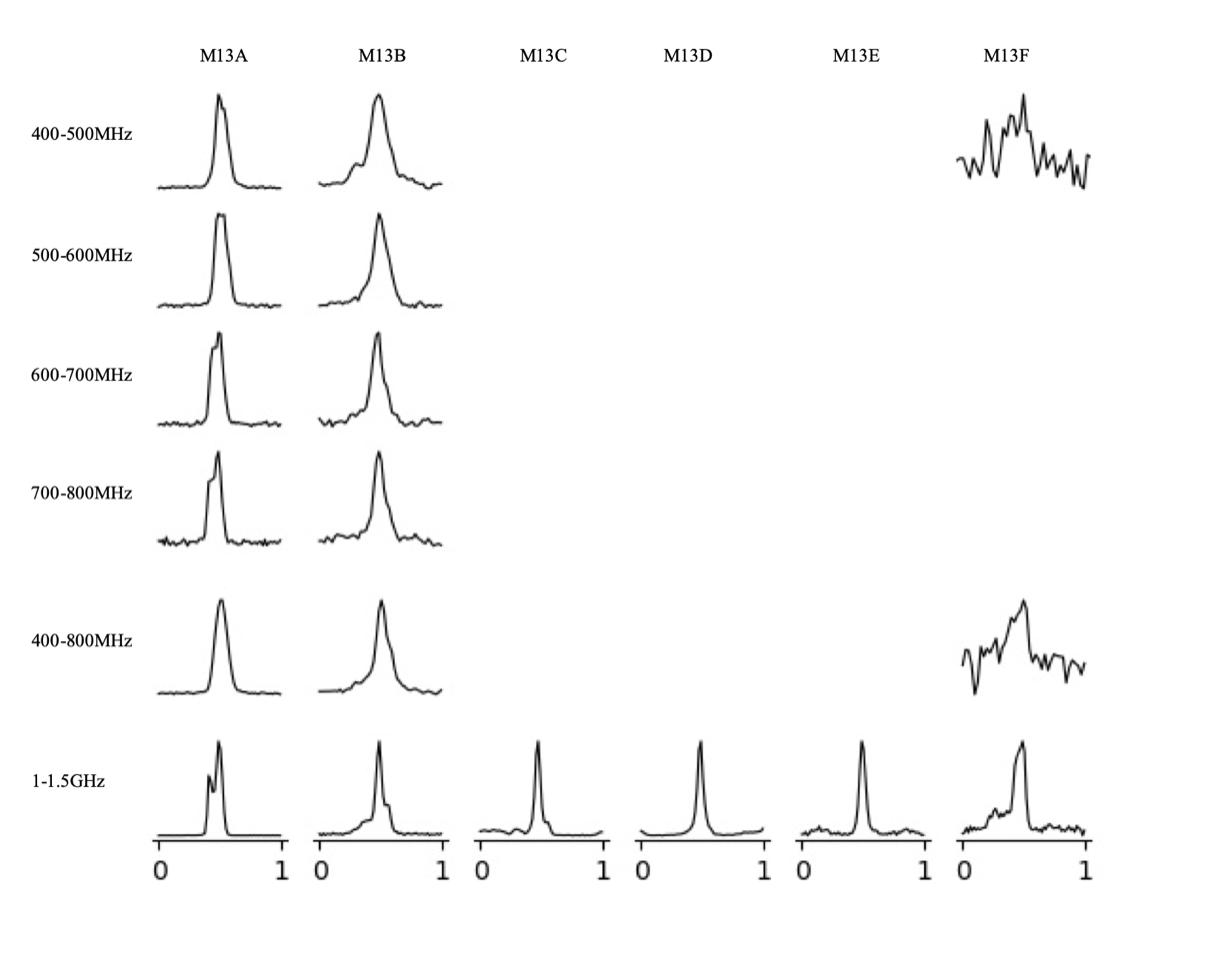}
\caption{Profiles of the 6 pulsars in M13 across different frequencies. X-axis of all the profiles are full rotations. The top four rows are the four 100\,MHz wide sub-bands of UWB 400$-$800\,MHz data and the fifth row is the sum of them. M13A, M13B and M13F were detected in low frequency bands. M13F was discovered in 400$-$800\,MHz, while it is only seen in 400$-$500\,MHz sub-band. \label{tab:knownpsr_profile}}
\end{figure*}

\subsection{Discovery of M13F and timing analysis}

The new pulsar, with a spin period of approximately 3.00\,ms, was first detected in the 400$-$800\,MHz UWB data and a summary of the detection can be found in Figure \ref{tab:searching_result}. Here the signal was seen across the whole frequency band, and detected at a DM of 30.4\,pc cm$^{-3}$ which is close to the average DM value of the known pulsars in M13. The P-Pdot diagram in the bottom-right panel shows that the signal has a significant period derivative, indicating the existence of acceleration to the pulsar most likely due to a binary companion. The 250$-$400\,MHz UWB data were also searched following the same approach and also with respect to the detection parameters from the discovery in the upper band, but the signal-to-noise (S/N) ratio was too low to confirm the presence of the signal at this frequency. The discovery was later confirmed with follow-up observations using the 19-beam at L-band. 19 out of the 24 follow-up observations detected M13F over the duration of approximately one year. The non-detections are most likely attributable to interstellar scintillation. 

\begin{figure*}[ht!]
\epsscale{1.3}
\plotone{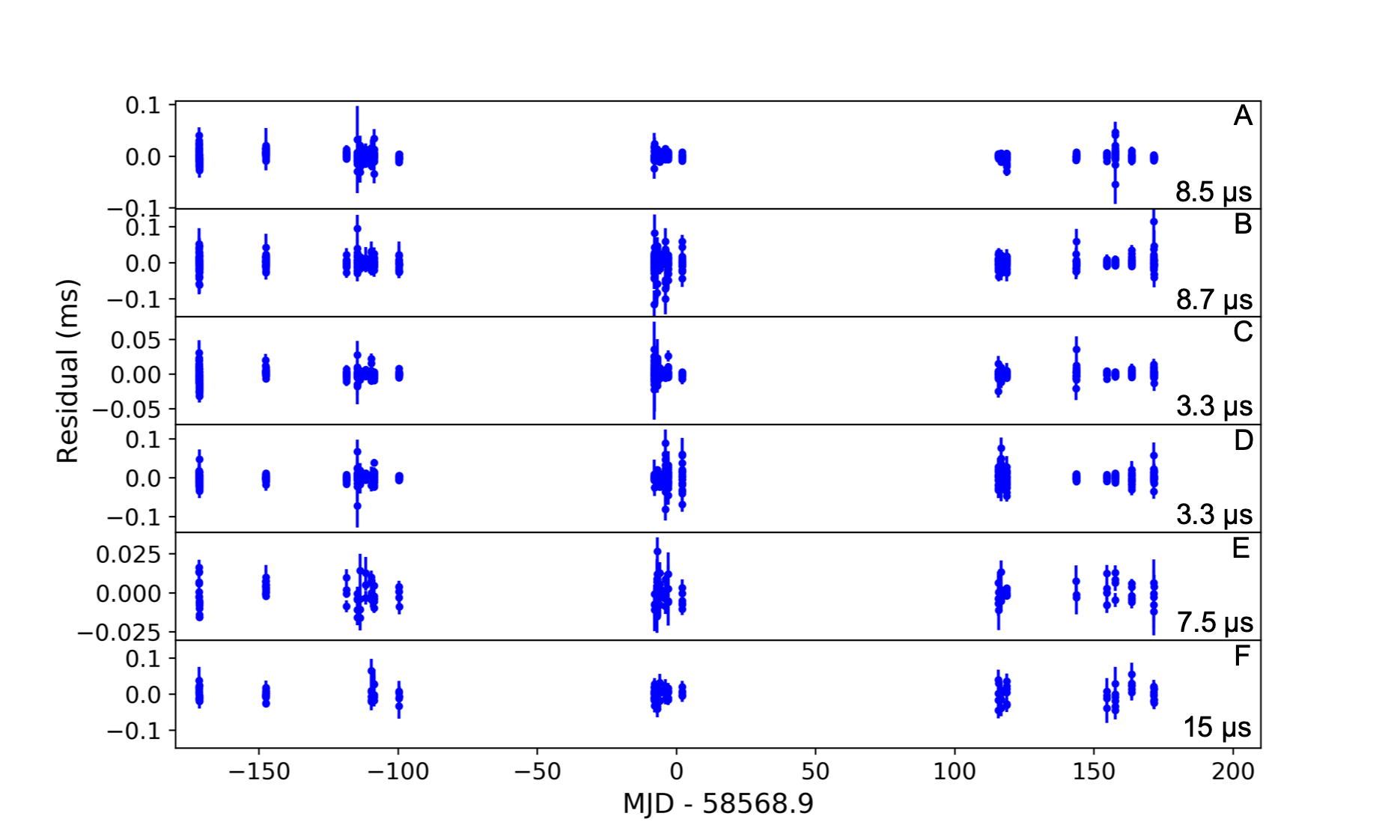}
\caption{The timing residuals from the best fit timing models presented in Table \ref{tab:timing_solution} for the six pulsars (as indicated in each panel). The values at the bottom-right of each panel are the RMS values for the residuals. All the data points are from the 19-beam data at L-band.\label{tab:residuals}}
\end{figure*}

\begin{figure*}[ht!]
\epsscale{1.3}
\plotone{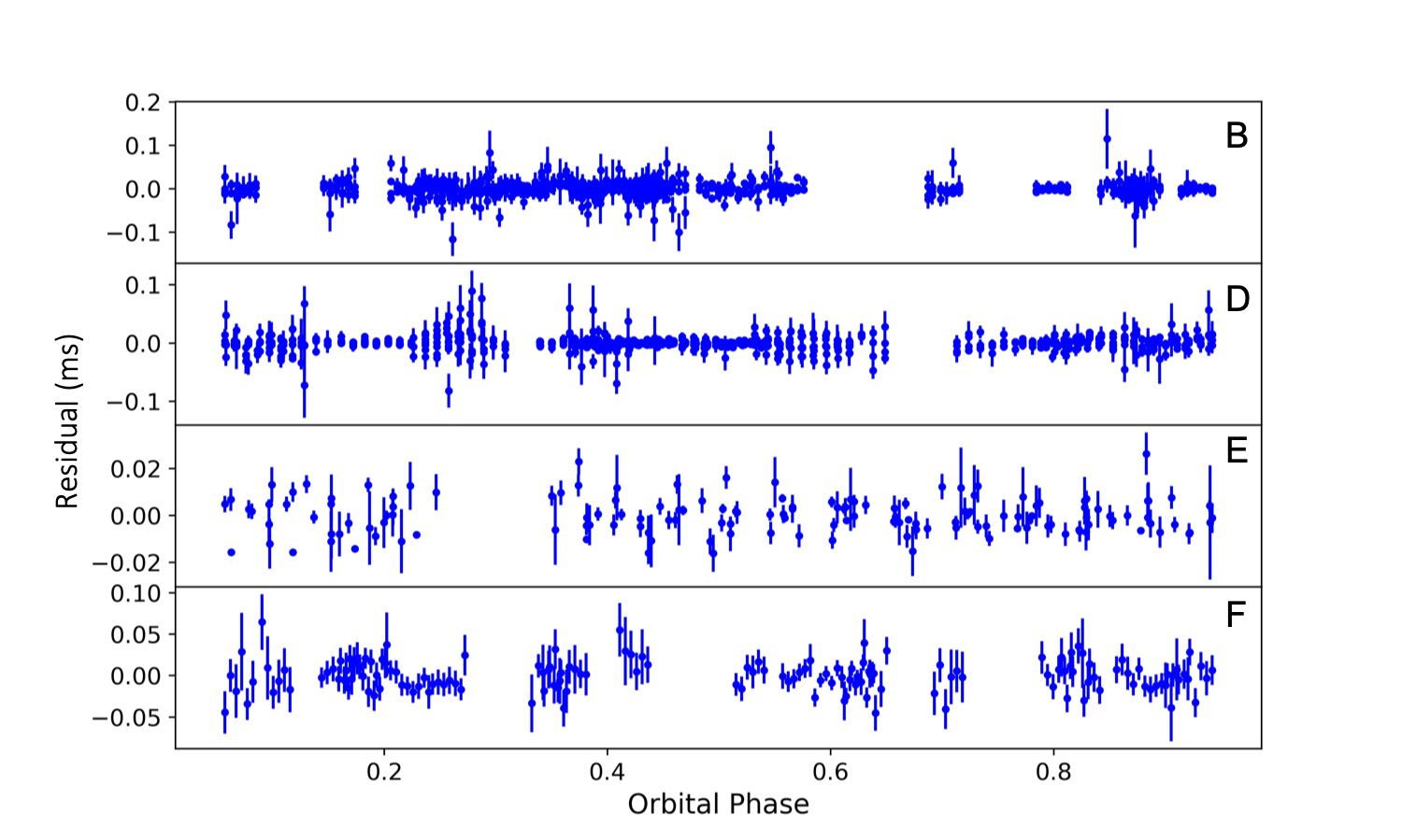}
\caption{Timing residuals as a function of orbital phase for the binary pulsars M13B, M13D, M13E and M13F. All the data points near the eclipsing phase of M13E are removed. \label{tab:orbital}}

\end{figure*}

In order to derive the orbital parameters of the system, for each epoch the raw search mode data was first folded into sub-integrations with 1\,min length each, using the \textsc{DSPSR} package \citep{2011vanStraten} and the ephemerides derived from the search results of the same data. From each sub-integration we generated its time-of-arrival (ToAs) using the {\tt pat} routine from the \textsc{PSRCHIVE} package. A plugin, {\tt stridefit2}, from the \textsc{TEMPO2} package \citep{Hobbs2006} was used to generate a series of measurements of spin period, which was used in the \textsc{fitorbit}\footnote{Software developed by A. G. Lyne to fit for binary pulsar parameters.} software package to fit for the binary parameters. We used the measurements from \textsc{fitorbit} as an initial guess and carried out a timing analysis of the ToAs with \textsc{tempo2} to obtain more accurate measurements of the timing parameters. These procedures were iterated a few times until the final coherent timing solution was converged to. DM was measured with \textsc{tempo2} using TOAs from multi-frequency sub-bands. Since a small eccentricity was shown in the initial timing solution, we thus use binary model ELL1 \citep{Lange2001} for M13F and the other binary pulsars in M13 in order to break degeneracy between orbital parameters. In Table~\ref{tab:timing_solution} we present the position, rotational and binary parameters of M13F. In Figure~\ref{tab:residuals} and Figure~\ref{tab:orbital}, the timing residuals are shown as functions of time and orbital phase. \\

\subsection{Detection and timing of known MSPs}

Five pulsars have been discovered in previous observations of M13. The timing solution of M13A was published in \citet{Kulkarni1991}, but hasn't been updated afterwards. There are no published timing solutions for M13B to E. 
To detect those pulsars, we first attempted to fold our data using the ephemeris from the ATNF pulsar catalogue\footnote{https://www.atnf.csiro.au/research/pulsar/psrcat/} \citep{Manchester2005}. However, we found those ephemeris are not accurate enough to keep phase coherence in our data, except for M13A. Thus, to detect M13B, M13C, M13D and M13E for each epoch observation, we performed a search for their periodic signal and then folded the data using the ephemerides from the search. Once the detections were made, we then followed the procedures as described in the previous subsection to obtain an accurate timing solution for all five pulsars. 

\begin{figure}[ht!]
\epsscale{1.3}
\plotone{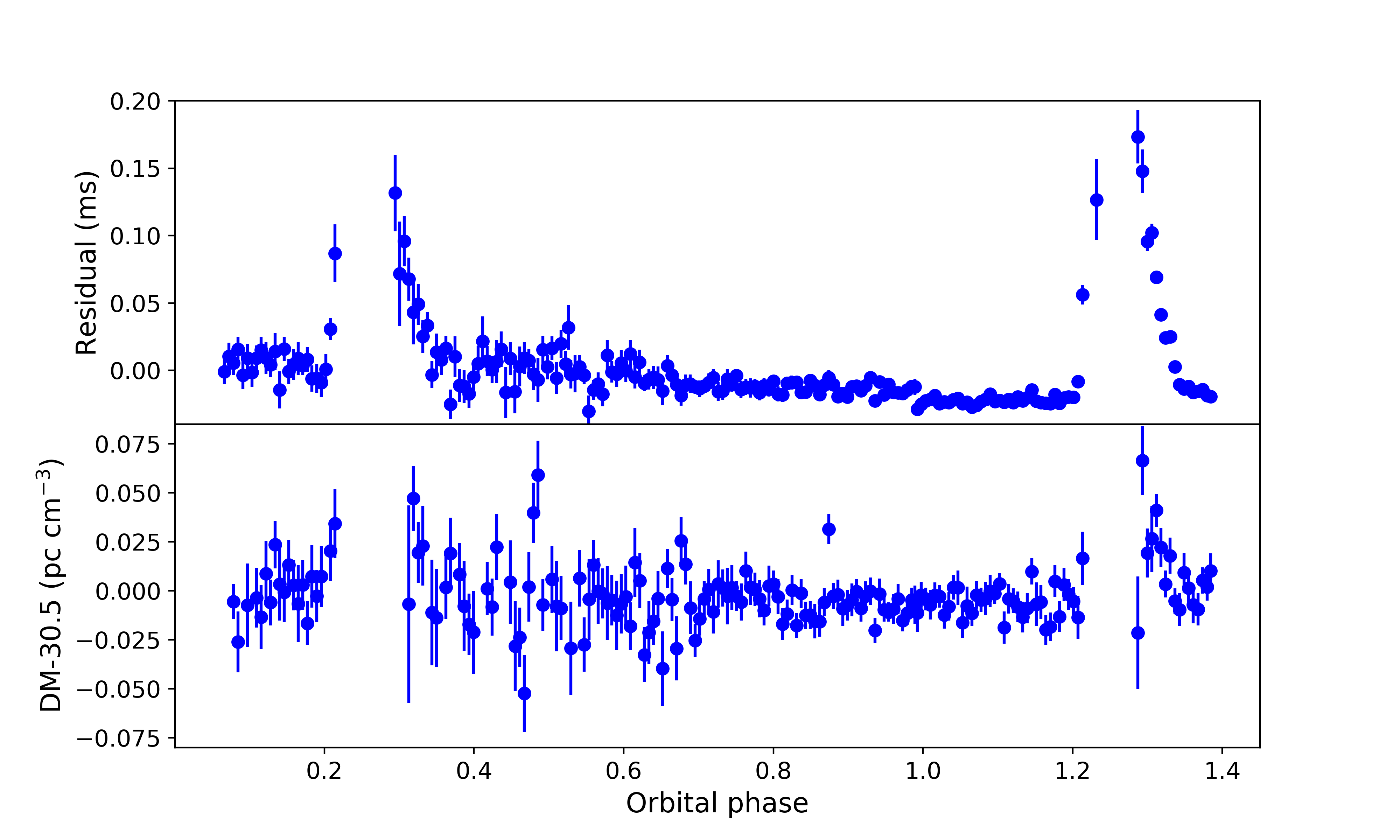}
\caption{Two eclipsing events of M13E. The data is from the observation 20181006. The integration time is 228\,min which is close to one and a half periods. The upper panel is residual as a function of orbital phase. The lower panel is DM variation.  \label{tab:eclipse}}
\end{figure}

As shown in Table \ref{tab:observations_table}, 
M13A to C were detected in all 19-beam observations. M13 D and E were detected in 92\% and 83\% of the 19-beam observations respectively. Only M13A and M13B were detected in the two sub-bands (250$-$400\,MHz and 400$-$800\,MHz) of the UWB data. Timing solutions of these pulsars are summarized in Table~\ref{tab:timing_solution}. M13A and M13C are the only isolated pulsars known in this cluster, the periods of which are 10.38\,ms and 3.72\,ms respectively. M13B is a 3.53\,ms MSP in a binary system with orbital period of 1.3 days. M13B has the largest position offset among the six pulsars in this cluster so far, which places it at the edge of the half-light radius of M13 (see Figure~\ref{tab:position}). M13B has the smallest DM (see Table~\ref{tab:timing_solution}) among the 6 pulsars in this cluster, this suggest that this pulsar is located on the nearside of the cluster, or it may be that the contribution to the gas internal to the cluster is less here. M13D is also a binary MSP with spin period and orbital period of 3.12\,ms and 0.6 days respectively. M13E has the fastest rotational period in the cluster and is closest to the center of the cluster. M13E displays eclipses near superior conjunction (at orbital phase 0.25). The eclipse event lasts for approximately 15\,mins which is about 10 $\%$ of the orbital period. Figure~\ref{tab:eclipse} shows the timing residuals and DM variation during the two eclipsing events of M13E. The DM variation is measured with 1-min sub-integrations using the frequency-resolved template-matching method developed in \citet{Liu2014}. The peaks in the plot, i.e., the delay to the arrival time of pulsar signal, are caused by increased DM which is contributed by the material outflowing from the companion star. It can be seen that during the eclipses the DM was increased by at least 0.05\,pc~cm$^{-3}$.

The profiles at different frequency bands of all the known pulsars can be found in Figure~\ref{tab:knownpsr_profile}. We only detected M13A, M13B and M13F in the frequency band 400$-$800\,MHz. We divided this band into 4 sub-bands, 400$-$500\,MHz, 500$-$600\,MHz, 600$-$700\,MHz and 700$-$800\,MHz respectively, M13A and M13B were seen in the 4 sub-bands and showed profile variation across the frequency bands (see Section below), while M13F was only seen in the sub-band 400$-$500\,MHz. 


\begin{table*}
\caption{Timing parameters for the six pulsars in M13, as obtained from fitting the observed ToAs with {\sc TEMPO2}. M13A and M13C are isolated pulsars. M13B, M13D, M13E and M13F are in binary systems. The companion masses are calculated assuming pulsar mass of 1.4\,M$_{\odot}$.}
\label{tab:timing_solution} 
\begin{center}{\scriptsize
\setlength{\tabcolsep}{6pt}
\renewcommand{\arraystretch}{1.3}
\begin{tabular}{l c c c}
\hline
Pulsar &  M13 A & M13 B & M13 C   \\
\hline\hline

Right Ascension, $\alpha$ (J2000)                                     \dotfill &   16:41:40.87019(5)            &    16:41:40.39144(7)            &   16:41:41.00748(3)   \\
Declination, $\delta$ (J2000)                                         \dotfill &   +36:27:14.9788(4)          &   +36:25:58.4880(5)         &   +36:27:02.7438(2)       \\
Spin Frequency, $f$ (Hz)                                        \dotfill &   96.362234567(3)       &      283.44094318360(2)      &     268.66686619507(6)    \\
First Spin Frequency derivative, $\dot{f}$ ($10^{-15}$ Hz s$^{-1}$)                \dotfill &   0.675(3) &   0.012(10)  &   $-$0.089(5) \\
Start of Timing Data (MJD)                                            \dotfill &   58396.28     &   58396.28           &   58396.28          \\
End of Timing Data (MJD)                                              \dotfill &  58741.46      &  58741.46                                                      &   58741.46   \\
Dispersion Measure, DM (pc cm$^{-3}$)                                 \dotfill &   30.4386(5)                                                              &   29.4456(6)                                                              &   30.1320(2)    \\
Number of ToAs                                                        \dotfill &  760                                                                 &   750                                        &   758   \\
Residuals RMS ($\mu$s)                    \dotfill &   8.506    &   8.735      &   3.307 \\

\hline
\multicolumn{4}{c}{Binary Parameters}  \\
\hline
 Binary Model                                                          \dotfill &   --                                                                     &   ELL1                                                          &   -- \\
Projected Semi-major Axis, $x_p$ (lt-s)                               \dotfill &   --                                                          &   1.388545(1)                                                            &   -- \\
Epoch of passage at Ascending Node, $T_\textrm{asc}$ (MJD)            \dotfill &   --                                                                     &  58561.7806094(1)                                                              &    -- \\
EPS1 \dotfill  & -- & 2(1)$\times 10^{-6}$ & -- \\
EPS2 \dotfill & -- & $-$2(1)$\times 10^{-6}$ & -- \\
Orbital Period, $P_{\rm b}$ (days)                                          \dotfill &   --         &    1.259112631(1)                                                       & -- \\

\hline
\multicolumn{4}{c}{Derived Parameters}  \\
\hline
Angular offset from centre in $\alpha$, $\theta_{\alpha}$ (arcsec) \dotfill & $-$0.36981 & $-$0.84856 & $-$0.23252  \\
Angular offset from centre in $\delta$, $\theta_{\delta}$ (arcsec) \dotfill & $-$20.5212 & $-$213.988 & $-$35.7562  \\
Spin Period, $P$ (ms)                                                  \dotfill &   10.3775094516(3)                                    &   3.5280718048989(3)                                     &   3.7220816029988(8)        \\
First Spin Period derivative, $\dot{P}$ ($10^{-21}$ s s$^{-1}$)                    \dotfill &   $-$7.3(3)   &   $-$0.1(1)  &   1.23(6)  \\
Orbital Eccentricity, $e$                                     \dotfill &   --         &   2(1)$\times 10^{-6}$    &   --  \\
Minimum companion mass, $M_{\rm c, min}$ (${\rm M}_\odot$)            \dotfill &   --    &   0.1605            &   --   \\
Median companion mass, $M_{\rm c, med}$ (${\rm M}_\odot$)             \dotfill &   --   &   0.1876                           &   --   \\
\hline

\hline
Pulsar  & M13 D & M13 E & M13 F  \\
\hline\hline
Right Ascension, $\alpha$ (J2000)                                     \dotfill &   16:41:42.395232(7)                                                       &   16:41:42.0221(1)                     &         16:41:44.6058(3)                                              \\
Declination, $\delta$ (J2000)                                         \dotfill &   +36:27:28.2021(3)                                                       &   +36:27:34.9676(9)                                &     +36:28:16.003（4）                       \\
Spin Frequency, $f$ (Hz)                                        \dotfill &   320.6886956910(2)                                                     &   402.0937023406(4)                                 &      332.9448049492(9)     \\
First Spin Frequency derivative, $\dot{f}$ ($10^{-15}$ Hz s$^{-1}$) \dotfill &   2.42(1) &   $-$2.82(2) &  $-$1.55(6) \\
Start of Timing Data (MJD)                                            \dotfill &   58397.29                                                             &   58396.29                                           &  58397.29  \\
End of Timing Data (MJD)                                              \dotfill &   58740.46                                                           &   58741.45                                         &   58740.46  \\
Dispersion Measure, DM (pc cm$^{-3}$)                                 \dotfill &   30.451(3)                                                               &   30.54(2)                                                            &  30.366(4)     \\
Number of ToAs                                                        \dotfill &   750                                                                    &   147                                                                 &      156                                        \\
Residuals RMS ($\mu$s)                                                \dotfill &   3.345                                                                  &   7.46                                                                  &  15.07                         \\

\hline
\multicolumn{4}{c}{Binary Parameters}  \\
\hline
Binary Model                                                          \dotfill &   ELL1                                                                   &   ELL1                                                                   &   ELL1                    \\
Projected Semi-major Axis, $x_p$ (lt-s)                               \dotfill &   0.9243183(4)                      &   0.035869(1)                      &        1.251702(3)          \\
Epoch of passage at Ascending Node, $T_\textrm{asc}$ (MJD)            \dotfill &   58397.38429971(5)                                                       &    58397.27621667(1)                                                       &  58398.0011780(7)             \\
EPS1  \dotfill  & 0.0001578(9) & 6(6)$\times 10^{-5}$ & 5(4)$\times 10^{-6}$ \\
EPS2  \dotfill  & 0.0005508(5) & -0.00011(5) & -2(4)$\times 10^{-6}$ \\
Orbital Period, $P_{\rm b}$ (days)                                          \dotfill &   0.5914408990(1)                                                         &   0.11261745834(4)                                    &       1.378005120(6)                   \\
\hline
\multicolumn{4}{c}{Derived Parameters}  \\
\hline
Angular offset from centre in $\alpha$, $\theta_{\alpha}$ (arcsec) \dotfill & +1.15523 & +0.782 & 3.3658  \\
Angular offset from centre in $\delta$, $\theta_{\delta}$ (arcsec) \dotfill & $-$7.2979 & $-$0.532 & $-$19.497\\
Spin Period, $P$ (ms)                                                  \dotfill &   3.118288899599(2)                                    &   2.486982497311(2)                                    &    3.003500835979(8)        \\
First Spin Period derivative, $\dot{P}$ ($10^{-20}$ s s$^{-1}$)                    \dotfill &   $-$2.36(1)  &   1.75(2)                                             &  1.40(6)   \\
Orbital Eccentricity, $e$                                             \dotfill & 0.0005730(6) &0.00012(5) & 5(4)$\times 10^{-6}$ \\
Minimum companion mass, $M_{\rm c, min}$ (${\rm M}_\odot$)            \dotfill &   0.1782                                                               &   0.01942                                                              &    0.1347              \\
Median companion mass, $M_{\rm c, med}$ (${\rm M}_\odot$)             \dotfill &   0.2085                                                               &   0.0225 &   0.1571   \\
\hline
\end{tabular} }
\end{center} 
\end{table*}

\section{Discussion} \label{sec:discuss}

\subsection{Properties of M13F} \label{subsec:M13}
M13F has the second shortest spin period and the longest orbital period among the six pulsars that have been discovered in M13 (see Table~\ref{tab:timing_solution}). The spin period derivative of M13F is 1.4 $\times$ 10$^{-20}$ , which is a typical value of MSPs in GCs\footnote{http://www.naic.edu/~pfreire/GCpsr.html}. The position offset of M13F (Figure~\ref{tab:position}) provides evidence that M13F is located at the edge of the cluster core. The median companion mass of M13F is 0.13\,M$_{\odot}$, which indicates that the companion is likely to be a white dwarf. The eccentricity of the system 5$\times$10$^{-6}$ is small compared to the ``normal" MSP$-$WD systems in 47 Tuc \citep{Freire2017}. This might be due to the low stellar density in this cluster which usually results in less encounter interactions. Encounter interactions were shown to be the main cause for eccentricities of low-mass binary MSPs in GCs which are supposed to be born with $e\approx10^{-6}-10^{-3}$ \citep{Rasio&Heggie1995,Heggie1996}. \\

\subsection{Profile evolution of M13A and M13B in frequency}

The profile of M13A shows two components, the leading component decreases in amplitude when compared to the trailing component as the observing frequency increases, indicating these two components have different spectral indices. Similar phenomena have been seen in other pulsars \citep[e.\,g.,][]{Dai2015}. M13B shows less apparent profile evolution in frequency, although the component at the leading edge fades out at frequencies above 500\,MHz. M13F also has multiple components as seen in Figure \ref{tab:knownpsr_profile}. While it is hard to tell the profile evolution of it due to the low S/N detection in the frequency range 400$-$800\,MHz.

\subsection{Sensitivity}
Of all the observations we have made, the longest one is 228 minutes, which corresponds to a minimum flux density of 0.4\,$\mu$Jy to a candidate with S/N$=$7. The sensitivity is calculated using the modified radiometer equation\citep{Lorimer2004}

\begin{equation}
S_{min} = \frac{\sigma\xi T_{sys}}{G \sqrt{n \Delta \nu T_{obs}}}\sqrt{\frac{W}{P-W}},\\    
\end{equation}
\\
where $\sigma$ represents the minimum S/N of a candidates.
Imperfections arise due to digitisation of the signal and other effects are incorporated in $\xi$ \citep{Lorimer2004}, and we use $\xi$ $=$ 1.0 as our data is sampled in 8-bit. $T_{sys}$ is the equivalent temperature of the observing system and the sky, which is 35\,K for the FAST 19$-$beam receiver. G is the gain of telescope and we use the value of 16 K/Jy \citep{Jiang2019}. We use two polarisations and so n=2. $\Delta \nu$ is the bandwidth and is effectively 400\,MHz in our L-band data. $T_{obs}$ is the integration time, W and P are pulse width and period respectively and we take 10\% for W/P. If we simply consider luminosity $L$ $=$ S$d^2$, where $S$ is flux density and $d$ is the distance of the cluster, we obtain a limit on luminosity with FAST of 19.9\,$\mu$Jy kpc$^{2}$. The weakest pulsar we have detected in GC M13 , M13F, has an L-band luminosity  of L $=$ 3.1\,mJy kpc$^{2}$, and we did not find any pulsar with luminosity lower than this value. We also probed different orbital phases of binaries and thus covered epochs where accelerations may have been small.  This may indicate that searches with higher sensitivity is needed if we are to find more pulsars in M13 and therefore require the SKA.\\

\subsection{Eccentricities}

\begin{figure}[ht!]
\epsscale{1.3}
\plotone{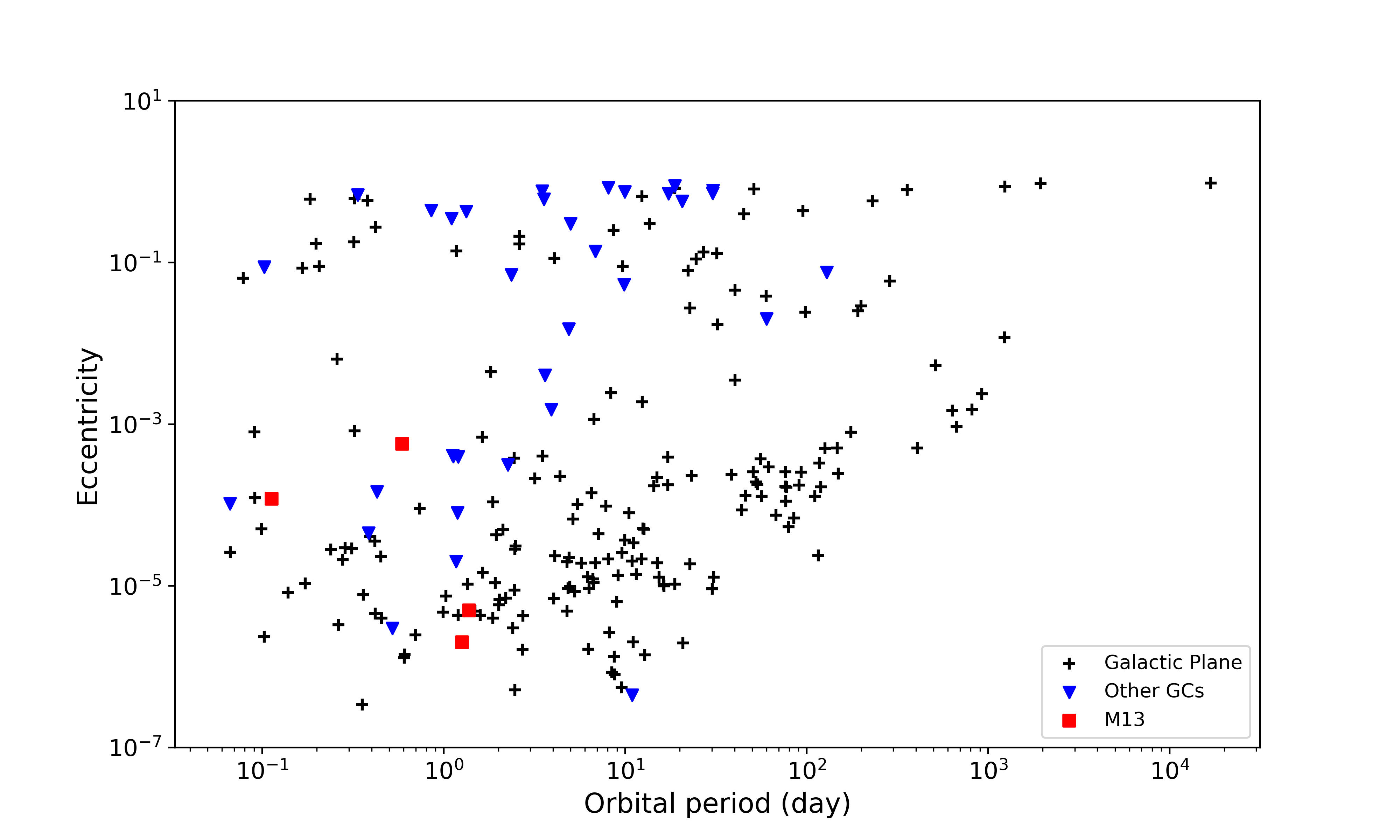}
\caption{ Eccentricity vs. orbital period of all binary pulsars. Black crosses are pulsars in the Galactic plane, green triangles are pulsars in GCs except M13 and red squares are binary pulsars in M13. The data used in this figure are from ATNF Pulsar Catalogue(https://www.atnf.csiro.au/research/pulsar/psrcat/)\citep{Manchester2005}. \label{tab:pb_ecc}}
\end{figure}

\begin{figure}[ht!]
\epsscale{1.2}
\plotone{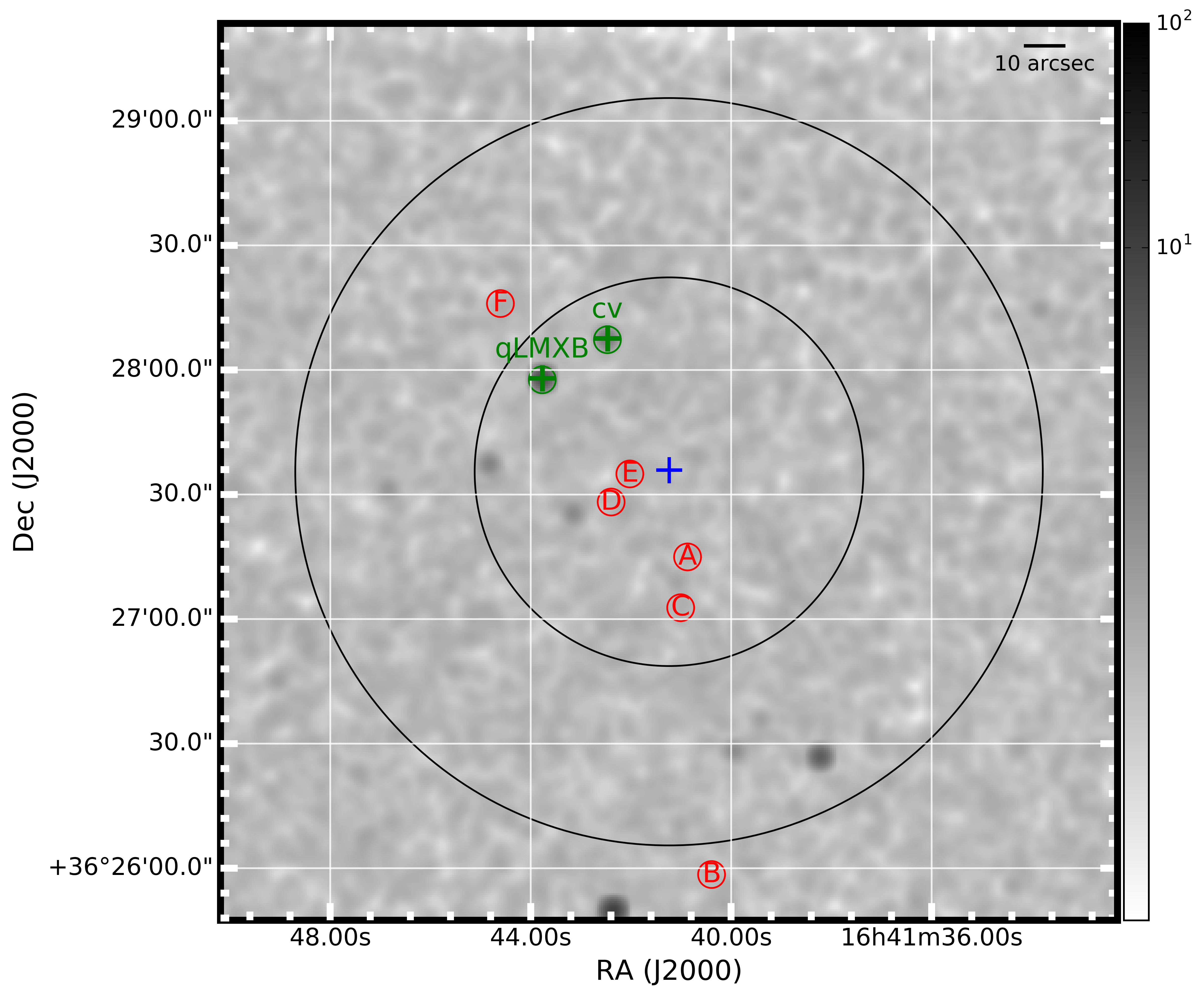}
\caption{Positions of the six pulsars in the GC M13, marked with red circles with letters. The green cross are a qLMXB and a cataclysmic variable (CV) which are also labeled in the plot. The centre of M13 is shown as a blue cross. The small circle in the middle is the core of M13 and the large circle is the half-light radius of M13. The background is the X-ray image of M13 from Chandra X-ray Observatory archive (OBsID 5436). Position errors of each source are too small to manifest on the plot. DM of pulsars in decending order is M13E, M13D, M13A, M13F, M13C and M13B.\label{tab:position}}
\end{figure}

Our results show that the eccenticities of pulsars in M13 are very small compared to what is found in other GCs. As plotted in Figure \ref{tab:pb_ecc}, most GC binary pulsars have relatively large eccentricities, while the eccentricities of pulsars in M13 are close to those in the Galactic plane. Excluding those pulsars in M13 only a small fraction of pulsars in GCs have eccentricities below ~10$^{-3}$ . The majority of this fraction are from the non-core-collapse cluster 47 Tuc \citep{Howell2000} and one is located in the cluster M62 which has an intermediate interaction rate per binary \citep{Verbunt2014}.  
This is in agreement with the prediction that more low-eccentricity pulsars are found in the clusters with very low interaction rate per binary \citep{Rasio&Heggie1995,Heggie1996}.
The eccentricities and positions in Table \ref{tab:timing_solution} and Figure \ref{tab:position}  seem to confirm that eccentricity is related to the stellar density around the pulsar. The pulsars in M13 close to the core have relatively large eccentricities, like M13D and M13E. While M13B has a small eccentricity and is located further from the core. The other pulsar with a large offset from the core is M13F, and it also has a small eccentricity. 

\subsection{Properties of pulsars in M13}
In \citet{Verbunt2014} the encounter rate for a single binary, $\gamma$, is used to characterize the difference between pulsar populations in GCs. In clusters with a low $\gamma$, a binary will have a relatively long life.
In such low-density environments, binaries will evolve undisturbed, and result in a MSP population similar to that observed in the Galaxy, where binary MSPs with low-mass WD and low-eccentricity significantly outnumber isolated pulsars.
M13 has the smallest $\gamma$ among the 14 GCs listed in \citet{Verbunt2014}. The number of discovered binary pulsars is twice more than that of isolated pulsars in M13, thus the result is well explained by the above assumption.  
A low $\gamma$ GC allow X-ray binaries to live long enough, so that the neutron star can be spun up to short periods \citep{Sigurdsson&Phinney1995}. The fact that the periods of all the six discovered pulsars in M13 are shorter than 0.015\,s is consistent with this expectation.\\

\begin{figure}[ht!]
\epsscale{1.3}
\plotone{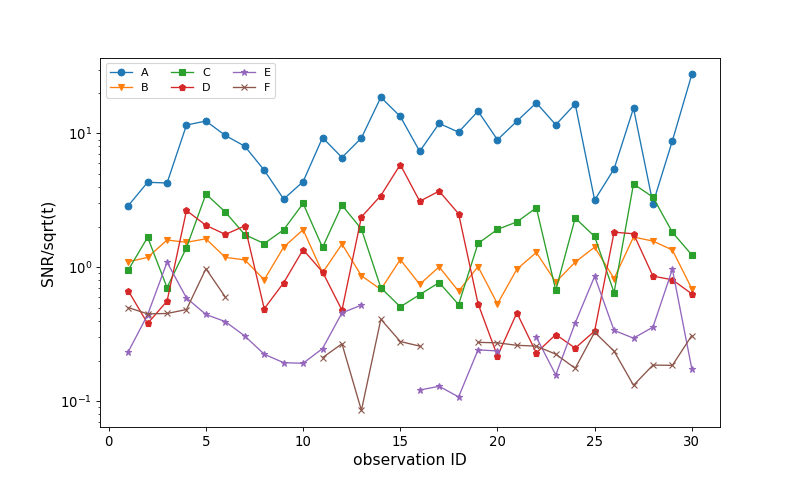}
\caption{S/N variations of the six pulsars in M13. x-axis is presented in observation ID, long observations are marked with several observation IDs in order that data in each observation ID have similar integration time. y-axis is S/N divided by $\sqrt{t}$. t is integration time of data in each observation ID.  \label{tab:SNR}}
\end{figure}

The interstellar medium (ISM) along the line of sight affects the pulsar signal which manifests as density variation \citep{Lorimer2004}. one of our initial thoughts is that if the pulsars are spatially close enough, the turbulence of the ISM along their signal propagation might turbulent in a similar way, which might lead to correlations between flux density variations of these pulsars. M13 has a DM value of approximately $30$\,cm$^{-3}$ pc and its pulsars have been seen to exhibit significant variations in flux density due to interstellar scintillation. In Figure~\ref{tab:SNR}, for each pulsar we plot the S/Ns per unit integration time from each observation, and calculate the correlation coefficient of each pair. No significant correlation is found between any pulsar pair. This indicates that the distance of each pulsar pair is not small enough so that the pulsars can scintillates simultaneously. \\
The difference between the largest and smallest DM value ($\delta$DM) in M13 pulsars is 1.1\,cm$^{-3}$ pc, this is the sixth smallest $\delta$DM among all the clusters that have been detected with more than one pulsar, the other 5 clusters with $\delta$DM smaller than M13 are 47 Tuc, M30, M3, M5 and NGC 6749, which are all relative low stellar density clusters. The ratio of $\delta$DM and average DM in M13 is 3.6\%, which is in agreement with the value around 3\% in 47 Tuc \citep{freire2001}. The contribution from the turbulence of Galatic electron column density to variations in DM is at the level of 0.05\,cm$^{-3}$ pc \citep{nordgren1992}. The difference we observed is 20 times larger than this value. Such DM difference could be caused by the gas distribution within the cluster. If a considerable population of pulsars can be discovered in M13, the gravitational field can be constrained by the DMs, positions, period derivatives and proper motions of pulsars in 3 dimensions \citep{freire2001,freire2001b,Freire2017}. This will provide the possibility to reveal the existence of an intermediate mass black hole \cite[e.\,g.][]{kzltan2017,Prager2017,perera2018,Abbate(b)2019,Abbate(a)2019} in the center of M13. \\

Two of the four binaries in M13 have measured eccentricities, namely, they both have non-circular orbits. This might allow to measure the rate of periastron advance ($\dot{\omega}$) and a method in \citet{ozel2016} can be used to measure the mass of pulsars.\\

The companion star in a black widow system can be strongly heated on the side of facing to the pulsar and this heating pattern could be observed in optical light curve \citep[e.\,g.][]{callanan1995,Romani2015}. Considering the accurate position we derived for the black widow system M13E, we expect to see such phenomenon of the system. However no evidence shows there is a variable star at the same position of M13E \citep{kopacki2003,pietru2004}.

\section{Conclusion} \label{sec:conclusion }
We have used the FAST telescope to search for pulsars in the GC M13 at center frequencies of 600 \,MHz and 1250\,MHz. These searches are part of our pulsar search survey in GCs during the FAST commissioning and early science phase. This is the most sensitive searches for the GC M13 up to now. Our survey has discovered a new MSP in a binary system and measured timing solutions for all known pulsars in the GC M13. We find that all the discovered binary systems in M13 having low eccentricities compared to those typical GC pulsars and the eccentricities decrease with distance from the core.

\acknowledgments
We thank all the people that have made contribution to this work. This work is supported by the Open Project Program of the Key Laboratory of FAST, NAOC, Chinese Academy of Sciences, and the project of Chinese Academy of Science (CAS) and the Jodrell Bank Centre for Astrophysics (JBCA) collaboration and the Grant No. Qian Education Contract KY[2019]214. This work made use of the data from FAST (Five-hundred-meter Aperture Spherical radio Telescope). FAST is a Chinese national mega-science facility, operated by National Astronomical Observatories, Chinese Academy of Sciences.  L.\,Wang acknowledges the financial support by China Scholarship Council (CSC) and the Science and Technology Facilities Council (STFC) of the UK. BWS acknowledges funding from the European Research Council (ERC) under the European Union’s Horizon 2020 research and innovation programme (grant agreement No. 694745). K.~Liu acknowledges the financial support by the European Research Council for the ERC Synergy Grant BlackHoleCam (Grant No. 610058), the FAST FELLOWSHIP from Special Funding for Advanced Users budgeted and administrated by Center for Astronomical Mega-Science, Chinese Academy of Sciences (CAMS). K.~Liu and B.~Peng acknowledge the CAS-MPG LEGACY funding "Low-Frequency Gravitational Wave Astronomy and Gravitational Physics in Space".

\vspace{5mm}
\facilities{FAST}


\software{PRESTO \citep{Ransom2002},  
          DSPSR \citep{2011vanStraten}, 
          TEMPO2 \citep{Hobbs2006}
          }

\bibliography{tex}



\end{document}